\documentclass[12pt,a4paper]{article}
\usepackage{t1enc}
\usepackage[dvips,final]{graphicx} 
\usepackage{tabularx} 
\usepackage{amsfonts}
\usepackage{amssymb}
\usepackage{amsmath}
\usepackage{epsfig}
\usepackage{epsf}
\usepackage{psfrag}

\def\du{\unskip\smash{\lower 1.4ex \hbox{\char34}}\kern-.2ex}
\def\hu{\kern-.2ex\hbox{\char92}}

\newcommand{\bdis}{\begin{displaymath}}
\newcommand{\edis}{\end{displaymath}}
\newcommand{\be}{\begin{equation}}
\newcommand{\ee}{\end{equation}}

\newcommand{\mcal}{\mathcal}

\newcommand{\noi}{\noindent}

\newtheorem{pr*}[thm]{*}

\begin{document}
\baselineskip=7mm
\newpage

\title{False vacuum decay with gravity in a critical case}
\author{Michal Demetrian \\
Comenius University \\
Mlynska Dolina M 105 Bratislava 4 \\
842 25 Slovak Republic}
\maketitle

\abstract{The vacuum decay in a de Sitter universe is
studied within semiclassical approximation for the class of effective inflaton
potentials whose curvature at the top is close to a critical value.
By comparing the actions of the Hawking - Moss instanton and the Coleman - de Luccia instanton(s) the mode of
vacuum decay is determined. The case when the fourth derivative of the effective potential at its top is less than
a critical value is discussed. }

\section{Introduction}

The idea of vacuum decay in a de Sitter universe (the transition of the inflaton field from false vacuum with positive
energy density to true vacuum with (almost) zero energy density caused by the quantum mechanical instability of
the false vacuum)
was developed by Coleman and de Luccia in \cite{cdl} and plays
an important role in the cosmological inflationary scenario. It is considered as a mechanism of transition
to a Friedman universe in old inflation \cite{guth} (in this scenario, the rapidly growing
bubbles of almost true vacuum are created in the sea of false vacuum; collisions of such bubbles were expected to
produce a Friedman universe, however it was established afterwards that there is not time enough for that)
and emerges also in the scenario of open inflation,\cite{bgt}, \cite{lm} (in this case, only single bubble is created and filled by the
configuration of inflaton that can evolve classically to the true vacuum; reheating is provided by decay
of inflaton particles into other particles via parametric resonance
\cite{lindekofman} at the stage when classical inflaton field oscillates around
the true vacuum).

We consider single scalar
field $\Phi$ with self-interaction given by the nonnegative function $V(\Phi)$ - effective potential - that has two
nondegenerate minima, one of them strictly positive (false vacuum) and second one equal to zero (true vacuum). These
vacua are supposed to be separated by a finite potential barrier. Let $V$  reach its local maximum $V_M$ at $\Phi_M$.
Furthermore, let us denote by $H(\Phi)$ the Hubble parameter corresponding to the energy-density $V(\Phi)$:
$H(\Phi)=\sqrt{8\pi V(\Phi)/3}$, especially $H_M=\sqrt{8\pi V_M/3}$. In order to study quantum transition of the
inflaton, in fact, one does not need to have the potential described above, namely the potential may have no
local minima (vacua), since the existence of potential barrier is sufficient for this purpose.
Supposing $O(4)$ symmetry supplemented by the
regularity we get the following (Euclidean) equations of motion and boundary conditions for Euclidean version of
the scale parameter $a$ and the inflaton $\Phi$
\begin{eqnarray} \label{ee}
& &
a''=-C\left(\Phi'^2+V\right)a, \quad \Phi''+3\frac{a'}{a}\Phi'=V'_{\Phi} \\
& &
\label{eel2}
a(0)=0, \ a'(0)=1, \Phi'(0)=\Phi'\left(\tau_f\right)=0,
\end{eqnarray}
where the constant $C$ equals $8\pi/3$ and $\tau_f>0$ is defined by the equation $a(\tau_f)=0$.
For any suitable potential there exists a trivial solution of the above problem that reads
\be \label{hm_instanton}
\Phi_{HM}=\Phi_M,\ a_{HM}=H_M^{-1}\sin\left(H_M\tau\right), \quad \mbox{with} \quad \tau_f=\frac{\pi}{H_M}
\ee
and is called the Hawking-Moss instanton \cite{hm}. This instanton mediates
the vacuum decay in such a way that
the inflaton jumps up to the top of the barrier in the horizon-size domain and afterwards the inflaton leaves
(by quantum or thermal fluctuations) the
unstable equilibrium and evolves classically to the true vacuum. There are also trivial solutions
corresponding to inflaton lying in stable equilibria in the true and false vacuum, respectively.
However, under some additional conditions, the
problem (\ref{ee}-\ref{eel2})
has also nontrivial solutions (with variable $\Phi$) called Coleman - de Luccia (CdL) instantons
(or bounces) \cite{cdl}. Following the ideas of paper \cite{vj2} (see also \cite{tanaka} and recently
\cite{weinberg}) CdL instantons can be characterized by how many times the inflaton crosses the top of the barrier. We
talk about the CdL instanton of the $l$th order if the inflaton crosses the top $l$-times.
\\
The boundary conditions (\ref{eel2}) provide the action of a CdL instanton (that follows from the general
Einstein-Hilbert action) to be finite. The action is a very important quantity for an instanton since it determines
the probability of the vacuum decay per unit space-time volume in the form $\exp(-S)$ .
This quantity
can be transformed, according to \cite{vj2}, into the following simple form
\be \label{act0}
S=2\pi^2\int_0^{\tau_f}\left[\left(\frac{1}{2}\Phi'^2+V\right)a^2-\frac{1}{C}
\left(aa'^2+1\right)\right]a{\rm d}\tau=-\frac{4\pi^2}{3C}\int_0^{\tau_f}a{\rm d}\tau .
\ee
It is easy to find that the action of the Hawking - Moss instanton is given by
\be \label{hmaction}
S_{HM}=-\frac{\pi}{H_M^2} .
\ee

\section{Near-to-limit CdL instanton of the first order and its action}

\begin{figure}[h]
\centering
\psfrag{a}
{$-\frac{V_M''}{H_M^2}$}
\psfrag{b}{$k_{num},\ k_{theor}$}
\includegraphics[]{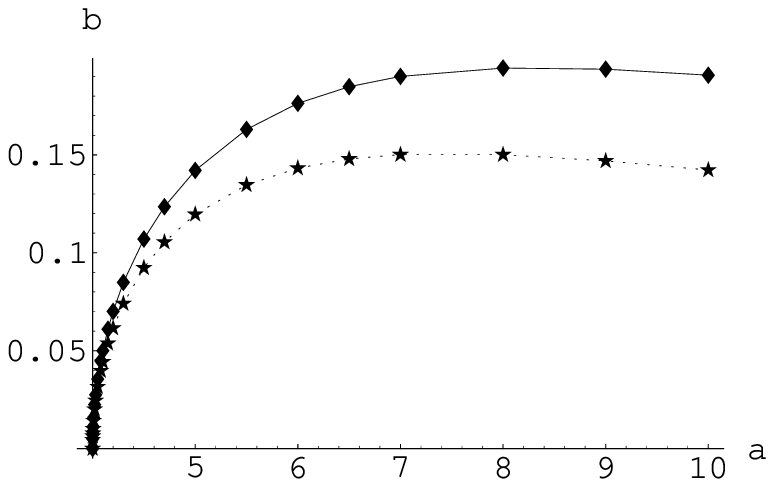}
\psfrag{a}
{$-\frac{V_M''}{H_M^2}$}
\psfrag{b}{$H_M^2w_{4\ num},\ H_M^2w_{4\ theor}$}
\includegraphics[]{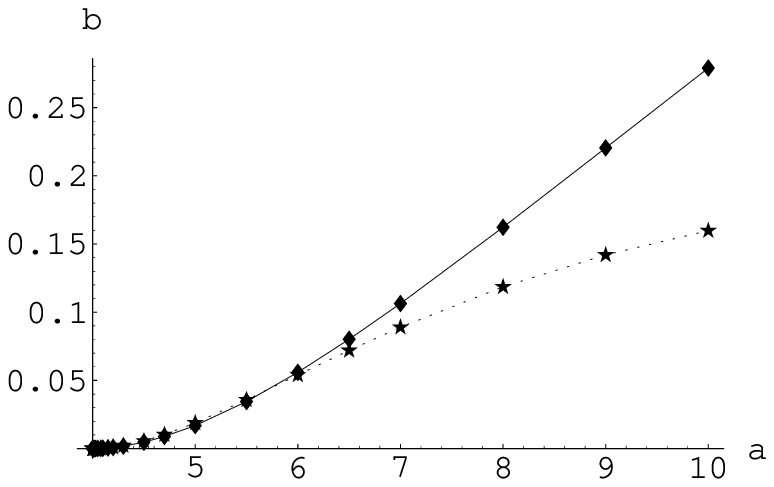}
\caption{The prediction of the analytical formula (\ref{k2quant}) for the instanton width in $\Phi$ direction
(lower, doted, line)
is compared with the numerical computations of this quantity in the left graph.
The
range of $-V_M''/H_M^2$ is taken $[4,10]$; at the value $4$ the CdL instanton of the first order appears and
at the value approximately $10$ the second order CdL instantons appear, \cite{vj1}, \cite{socdlja}.
Finally,
the right graph shows the theoretical dependence of the first-order CdL instanton action according to (\ref{w4})
(lower, doted, line) together with numerically obtained values of this quantity.}
\label{f1}
\end{figure}

As it was sketched in \cite{js} and finally proved in \cite{vj1} the CdL instanton necessarily exists for
potentials with $V_M''/H_M^2<-4$ and may exist if $V''/H^2<-4$ for some value of $\Phi$ in the potential barrier. If the
fraction $V_M''/H_M^2$ approaches one of the values $-l(l+3),\ l=1,2,3 \dots$, the near-to-limit CdL instanton (that
approaches necessarily existing Hawking-Moss instanton) may exist. If $V_M''/H_M^2<-l(l+3)$ then the instanton
of $l$th order necessarily exists. Our task is to compute the difference between the action of the near-to-limit
CdL instanton of the first order and the action of the related Hawking - Moss instanton. This task has been considered in
\cite{vj2}, but our treatment will be different.
By making use of the re-scaled Euclidean time
$x=H_M\tau$ and the shifted inflaton field $y=y(x)=\Phi(\tau(x))-\Phi_M$ we rewrite equations (\ref{ee}) in to the form
\be \label{ee1}
a''=-C\left(y'^2+\frac{V}{H^2_M}\right)a, \quad y''+3\frac{a'}{a}y'=\frac{V'_y}{H_M^2}.
\ee
Since now the prime denotes differentiation with respect to $x$. Introducing the expansion of the relevant
quantities, including the dimensionless Euclidean action
\bdis
\sigma=-\frac{3CH_M^2}{4\pi^2}S , \qquad \left(\sigma_{HM}=2\right)
\edis
into the series in the inflaton amplitude $k$
\be \label{k-exp}
y(x)=\sum k^nu_n(x), \ -\frac{V_M''}{H_M^2}=4+\sum k^n\Delta_n, \ a(x)=CH_M^{-1}\sum k^nv_n(x), \
\sigma=2+\sum k^nw_n
\ee
and expanding the potential into the powers of $y$
\bdis
V=V_M+\frac{1}{2}V_M''y^2+\frac{1}{6}V_M'''y^3+\frac{1}{24}V_M''''y^4+ \dots
\edis
we replace the nonlinear equations (\ref{ee1}) by the infinite system of linear equations
\begin{eqnarray} \label{linsys}
& &
u_n''(x)+3\frac{\cos(x)}{\sin(x)}u_n'(x)+4u_n(x)=\mcal{U}_n(x), \nonumber \\
& &
v_n''(x)+v_n(x)=\mcal{V}_n(x)\sin(x)
\end{eqnarray}
in which the functions $\mcal{U}_n$ and $\mcal{V}_n$ can be computed order by order if we know
the functions $u_{n-1},u_{n-2},\dots ,u_0$. Functions $u_n$ and $v_n$ are defined on the interval
$[0,x_f^{(n)}]$, where $x_f^{(n)}$ is defined as the point in which the scale factor $a$ computed up to the $n$th order
in $k$ vanishes. The value of the functions $v_n$ and $v_n'$ must vanish at $x=0$ (this
follows from (\ref{eel2})) and $u_n$ must be regular.
We know that
\bdis
u_0=0, \ v_0=\sin\left(x\right) \quad \mbox{and} \quad u_1=\cos(x) .
\edis
Furthermore, $V_M'=0$ implies that $v_1$ vanishes. The next nonzero term in $a$ is given by $v_2$ that must obey
\bdis
v_2''+v_2=-\frac{1}{4}\left[\sin(x)-3\sin(3x)\right].
\edis
The solution is
\be \label{v2}
v_2(x)=\frac{1}{4}\left[\frac{5}{8}\sin(x)+\frac{1}{2}x\cos(x)-\frac{3}{8}\sin(3x)\right] .
\ee
Solving equation $v_0(x)+k^2v_2(x)=0$ with the accuracy up to the order $k^2$ and
supposing the solution is close $x=\pi$, we find that the shifted right end-point is given by
\bdis
x_f^{(2)}=\pi-\frac{1}{8}C\pi k^2\equiv \pi+\delta^{(2)} .
\edis
Knowing $v_2$ we can compute the contribution of the order $k^2$
to the difference between the actions of CdL and Hawking - Moss instanton that is defined by eqs. (\ref{act0}) and
(\ref{k-exp}). The result is
\bdis
w_2=\int_0^{\pi}v_2(x){\rm d}x=0.
\edis
This means that we cannot distinguish between the action of
a near-to-limit CdL instanton and the related Hawking - Moss instanton in the second order of inflaton amplitude and
we must continue our computations. Equation for $u_2$ reads
\bdis
u_2''+3\frac{\cos(x)}{\sin(x)}u_2'+4u_2=\frac{1}{2}\frac{V_M'''}{H_M^2}\cos^2(x)
\edis
and its regular solution is
\be \label{u2}
u_2(x)=\frac{1}{24}\frac{V_M'''}{H_M^2}\left[1-2\cos^2(x)\right] .
\ee
Now, we can derive equation for $v_3$ and its solution
\be \label{v3}
v_3''+v_3=\frac{V_M'''}{48H_M^2}\left[2\sin(2x)-5\sin(4x)\right] \ \Rightarrow
v_3=-\frac{V_M'''}{72H_M^2}\left[\sin(2x)-\frac{1}{2}\sin(4x)\right] .
\ee
There is no shift of the right end point $x_f$ since $v_3(\pi)=0$,
and we can compute the $k^3$-contribution to the action as follows
\bdis
w_3=\int_0^\pi v_3(x){\rm d}x=0.
\edis
This result forces us to continue up to the fourth order in $k$. The shift of the right end point $x_f$
with respect to $\pi$ (which
is actually of the order $k^2$)
must be taken under consideration in the equation for $u_3$. Introducing new a independent variable
\bdis
X=\frac{\pi x}{\pi+\delta^{(2)}}=x\left(1+\frac{1}{8}Ck^2+o\left(k^2\right)\right)\equiv K x
\edis
we obtain supplementary formulas
\bdis
v_0\left( K^{-1}X\right)=\sin\left( X\right)-\frac{Ck^2}{8}X\cos\left( X\right),\
v_2\left( K^{-1}X\right)=v_2\left( X\right),
\edis
\bdis
v_0(x)+k^2v_2(x)=\sin\left( X\right)\left[1-\frac{Ck^2}{8}X\frac{\cos\left( X\right)}{\sin\left( X\right)}+
\frac{1}{4}Ck^2\left(1-\frac{3}{2}\cos^2\left( X\right)\right)\right] \equiv
\sin\left( X\right)+e\left( X\right)
\edis
and
\bdis
\frac{v_0'(x)+k^2v_2'(x)}{v_0(x)+k^2v_2(x)}=K\frac{\cos\left( X\right)+\frac{{\rm d}e\left( X\right)}{{\rm d}X}}
{\sin\left( X\right)+e\left( X\right)}=K
\left[\frac{\cos\left( X\right)}{\sin\left( X\right)}+\frac{3}{4}Ck^2\cos\left( X\right)\sin\left( X\right)\right]
\edis
which provides us with the equation for $u_3$ of the form
\begin{eqnarray*}
& &
\frac{{\rm d}^2u_3\left( X\right)}{{\rm d}X^2}+3\frac{\cos\left( X\right)}{\sin\left( X\right)}
\frac{{\rm d}u_3\left( X\right)}{{\rm d}X}-
\frac{V_M''}{H_M^2}u_3\left( X\right)=\\
& &
\left[ \frac{1}{K^2}\left( 4+\frac{V_M''}{H_M^2}\right)+\frac{13}{4}C+
\frac{1}{24}\left(\frac{V_M'''}{H_M^2}\right)^2\right]\cos\left( X\right)+
\left[\frac{1}{6}\frac{V_M''''}{H_M^2}-\frac{9}{4}C-
\frac{1}{12}\left(\frac{V_M'''}{H_M^2}\right)^2\right]\cos^3\left( X\right) \equiv \\
& &
A\cos\left( X\right)+B\cos^3\left( X\right).
\end{eqnarray*}
The only regular ($\frac{{\rm d}u_3(0)}{{\rm d}X}=\frac{{\rm d}u_3(\pi)}{{\rm d}X}=0$)
solution to this equation is given by the formula
\bdis
u_3\left( X\right)=\beta\cos^3\left( X\right)
\edis
with the constant $\beta$ to be determined from the system of linear equation
\bdis
6\beta=A,\ -14\beta=B\ \Rightarrow \
\beta=-\frac{1}{14}
\left\{\frac{1}{6}\frac{V_M''''}{H_M^2}-\frac{9}{4}C-\frac{1}{12}
\left(\frac{V_M'''}{H_M^2}\right)^2\right\} .
\edis
However, the fixation of $\beta$ is only a supplementary consequence of previous system of linear equations,
since their main purpose is to determine the value of $k^2$ as a function of $A$ and $B$.
Namely, we obtain from them the following
"quantization rule" for $k^2$ as a function of $4+V_M''/H_M^2$
\be \label{k2quant}
k^2=-\frac{4+\frac{V_M''}{H_M^2}}
{\frac{2}{7}\left[8C+\frac{1}{48}\left(\frac{V_M'''}{H_M^2}\right)^2+\frac{1}{4}\frac{V_M''''}{H_M^2}\right]} .
\ee
If the denominator of the fraction on the right hand side is positive then we have, for small
negative numerator a near-to-limit CdL instanton of the first order whose inflaton amplitude
is given by (\ref{k2quant}). We will return to the case when the denominator is negative later.
Now, let us concentrate
on computation of the action of this near-to-limit CdL instanton. Performing some
tedious algebra one derives equation for $v_4$ of the form
\bdis
v_4''+v_4=\left[\aleph_0+\aleph_2\cos^2(x)+\aleph_4\cos^4(x)\right]\sin(x) ,
\edis
where we have introduced the parameters
\begin{eqnarray*}
\aleph_0 & = &
-\frac{15}{16}C+\frac{1}{288}\left(\frac{V_M'''}{H_M^2}\right)^2 , \\
\aleph_2 & = &
\frac{159}{224}C-\frac{2}{21}\left(\frac{V_M'''}{H_M^2}\right)^2+\frac{3}{28}\frac{V_M''''}{H_M^2} \\
\aleph_4 & = &
\frac{27}{56}C+\frac{1}{7}\left(\frac{V_M'''}{H_M^2}\right)^2-\frac{9}{56}\frac{V_M''''}{H_M^2} .
\end{eqnarray*}
The solution is
\begin{eqnarray} \label{v4}
& &
v_4=\frac{1}{192}\left\{ -12\left(8\aleph_0+2\aleph_2+\aleph_4\right)x\cos(x)+ \right. \nonumber \\
& &
\left.
\sin(x)\left[ 96\aleph_0+36\aleph_2+23\aleph_4-2\left(6\aleph_2+5\aleph_4\right)\cos(2x)-2\aleph_4
\cos(4x)\right]\right\}
\end{eqnarray}
and with the help of it we finish the computations with a nonzero contribution to the action of a
surprisingly simple form
\be \label{w4}
\Delta S^{(4)}\equiv -\frac{4\pi^2k^4w_4}{3CH_M^2}=\frac{2\pi^2}{15}\frac{k^2}{H_M^2}
\left( 4+\frac{V_M''}{H_M^2}\right) .
\ee
Formula (\ref{w4}) tells us that a near-to-limit CdL instanton of the first order has, in the case
$V''_M/H_M^2<-4$, less action than the related Hawking - Moss instanton and therefore, if no other instantons exist,
it is the instanton governing the false vacuum decay. \\
Let us demonstrate the power of the formulas (\ref{k2quant}) and (\ref{w4}) on a concrete example. We will consider
the often mentioned quartic potential
\be \label{quarticpot}
V(\Phi)=\frac{1}{2}\Phi^2-\frac{1}{3}\delta\Phi^3+\frac{1}{4}\lambda\Phi^4 ,
\ee
where $\delta$ and $\lambda$ are supposed to be positive. The non-negativeness of the potential and the existence of
the false vacuum (the true vacuum is located at $\Phi=0$) require that the $\delta$ parameter belongs to the interval
$(\delta_m,\delta_M)$, where
\bdis
\delta_m=2\sqrt{\lambda},\ \delta_M=3\sqrt{\frac{\lambda}{2}}\approx 1.06 \delta_m .
\edis
The positions of the false vacuum ($\Phi_{fv}$) and of the top of the barrier are given by
\bdis
\Phi_{fv}=\frac{\delta}{2\lambda}+\sqrt{\frac{\delta^2}{4\lambda^2}-\frac{1}{\lambda}}=
\frac{\sqrt{1-Z^2}}{(1-Z)\sqrt{\lambda}}, \
\Phi_M=\frac{\delta}{2\lambda}-\sqrt{\frac{\delta^2}{4\lambda^2}-\frac{1}{\lambda}}=
\frac{\sqrt{1-Z^2}}{(1+Z)\sqrt{\lambda}} ,
\edis
where
\bdis
Z=\sqrt{1-\frac{4\lambda}{\delta^2}}, \quad Z\in\left[ 0,\frac{1}{3}\right] .
\edis
The potential (\ref{quarticpot}) in the $(Z,\lambda)$ parametrization has the form
\bdis
V=\frac{1}{2}\Phi^2-\frac{2}{3}\frac{\sqrt{\lambda}}{\sqrt{1-Z^2}}\Phi^3+\frac{1}{4}\lambda\Phi^4 .
\edis
We are interested in the quantities
\bdis
H_M^2\equiv \frac{8\pi}{3}V_M=\frac{2\pi}{9\lambda}\frac{(1-Z)(1+3Z)}{(1+Z)^2}, \quad
V_M''=-\frac{2Z}{1+Z} .
\edis
From these expressions it follows that the effective curvature of the potential at its top is given by
\be \label{efcurvq}
\frac{V_M''}{H_M^2}=-\frac{9\lambda}{\pi}\frac{Z(1+Z)}{(1-Z)(1+3Z)}
\ee
and is a monotonically increasing function of $Z$.
If we denote by $\lambda_S$ the value of $\lambda$ at which (at given $Z$) $V_M''/H_M^2=-4$, then
\bdis
\lambda_S=\frac{4\pi}{9}\frac{(1-Z)(1+3Z)}{Z(1+Z)} .
\edis
It will be useful to express the fractions $V'''_M/H_M^2$ and $V_M''''/H_M^2$ entering the formulae
(\ref{k2quant}) and (\ref{w4}) in terms of the parameters of the quartic potential. After some algebra one finds out
that
\bdis
\frac{V_M'''}{H_M^2}=-\frac{9\lambda^{3/2}}{2\pi}\frac{\left(1+Z\right)^2}{(1-Z)\sqrt{1-Z^2}}, \quad \mbox{and} \quad
\frac{V_M''''}{H_M^2}=\frac{27\lambda^2}{\pi}\frac{\left( 1+Z\right)^2}{(1-Z)(1+3Z)}
\edis
and by using the relation (\ref{efcurvq}) one gets the dependence of the fractions in question on the effective
curvature $V_M''/H_M^2$ of the potential at its top.
\be \label{frfo}
\frac{V_M'''}{H_M^2}=\frac{\sqrt{\pi}}{6}\left(\frac{1+Z}{1-Z}\right)^{1/2}\left(\frac{1}{Z}+3\right)^{3/2}
\left(-\frac{V_M''}{H_M^2}\right)^{3/2}, \
\frac{V_M''''}{H_M^2}=\frac{\pi}{3}\frac{(1-Z)(1+3Z)}{Z^2}\left(\frac{V_M''}{H_M^2}\right)^2 .
\ee
Now, we are ready to compare the predictions of the formulas (\ref{k2quant}) and (\ref{w4}) with the
numerical solutions of exact instanton equations (\ref{ee1}). In order to perform the numerical analysis of the
instanton equations we will fix the parameter $Z$ of the quartic potential to have the value
corresponding to central point
between the case when the false vacuum energy density $V_{fv}$ is negligible in comparison to
$V_M$ (this situation corresponds to the thin-wall approximation considered in \cite{cdl} and
is analyzed in part numerically in \cite{samuel}),
and the case when the false vacuum disappears. Since the energy density in false vacuum is given by
\bdis
V_{fv}=\frac{1}{12\lambda}\frac{(1+Z)(1-3Z)}{(1-Z)^2} ,
\edis
the fraction $V_{fv}/V_M$, that depends on $Z$ only, equals $1/2$ if
\bdis
\left(\frac{1+Z}{1-Z}\right)^3\frac{1-3Z}{1+3Z}=\frac{1}{2} \quad \mbox{with} \quad Z\in[0,1/3].
\edis
This equation determines $Z$ as
\be \label{Z_value}
Z\approx 0.278\ .
\ee
(For the values of $Z$ for which $V_{fv}/V_M\lessapprox 1/2$
the function $V_{fv}/V_M$ can be replaced with a good accuracy
by $-12Z+12/3$; this expression would give $Z=21/72\approx 0.292$.) We have solved numerically
the exact instanton equations with this choice of the parameter $Z$ and compared these results with the approximative
formulae (\ref{k2quant}) and (\ref{w4}), see figure \ref{f1}.

\section{CdL instanton(s) of the first order in the case with subcritical value of the fourth derivative of the
effective potential at its top}

Let us consider an effective potential which has,
for a suitable choice of parameters, such a shape that the denominator in the formula (\ref{k2quant})
is negative and at the same time it is possible to change continuously the sign of the nominator. If the sign of the term
$-4-V_M''/H_M^2$ is positive, then there must be at least one CdL instanton of the first order,
as discussed previously. But what
happens when we pass through zero to negative values of $-4-V_M''/H_M^2$, keeping
$\left[8C+\frac{1}{48}\left(\frac{V_M'''}{H_M^2}\right)^2+\frac{1}{4}\frac{V_M''''}{H_M^2}\right]$ a negative
constant? The formula (\ref{k2quant}) ensures that we have the near-to-limit CdL instantons in the
region with $-V_M/H_M^2$ (a little bit) less than $4$. By the continuity argument, this set of instantons
must be lined-up to the "overcritical" instantons existing for $-V_M''/H_M^2>4$.
In order to investigate the structure of the
instanton solutions it is helpful to use the method of representation of a
CdL instanton proposed by Tanaka in \cite{tanaka}. Let us consider the two dimensional "phase" plane
$(\Pi,\Phi)$, where $\Phi$ stands for some value (to be determined later) of the inflaton and $\Pi$ stands for some
value of the conjugated momentum $2\pi^2a^3\Phi'$. For a given $V$ we can start the evolution,
using to the Euclidean equations of motion (\ref{ee}), (\ref{eel2}) for $a$ and $\Phi$, with
any initial value $\Phi_i$ of the inflaton and we will come, in some finite Euclidean time $\bar{\tau}(\Phi_i)$,
to the point at which
$a$ reaches its maximum. Let $\Phi_i^+$ be an arbitrary value of $\Phi$ located to
the right of $\Phi_M$. Taking this $\Phi_i^+$
as the initial value for the system (\ref{ee}), (\ref{eel2}) we obtain the point
$(\bar{\Pi}^+,\bar{\Phi}^+)\equiv (\Pi(\bar{\tau}),\Phi(\bar{\tau}))$, and
varying $\Phi_i^+$ we can draw the curve
\bdis
\mcal{C}^+=\left\{ (\bar{\Pi}^+,\bar{\Phi}^+),\ \Phi_i^+>\Phi_M\right\} .
\edis
Analogically, varying the initial value of the inflaton $\Phi_i^-$ located to the left of $\Phi_M$ we construct the curve
$\mcal{C}^-$. Intersections of the curves $\mcal{C}^+$ and $\mcal{C}^-$ correspond to the CdL instantons.
(The curve $\mcal{C}^+$ does not intersect itself and the same holds for $\mcal{C}^-$).

\begin{figure}
\centering
\includegraphics[]{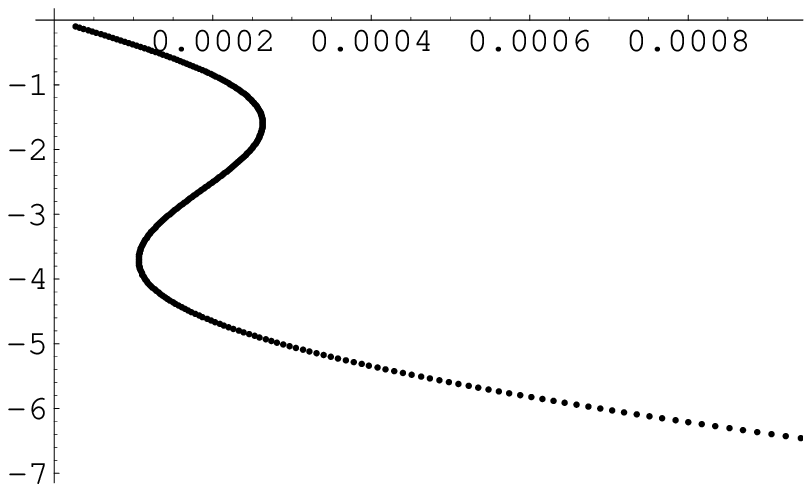}
\includegraphics[]{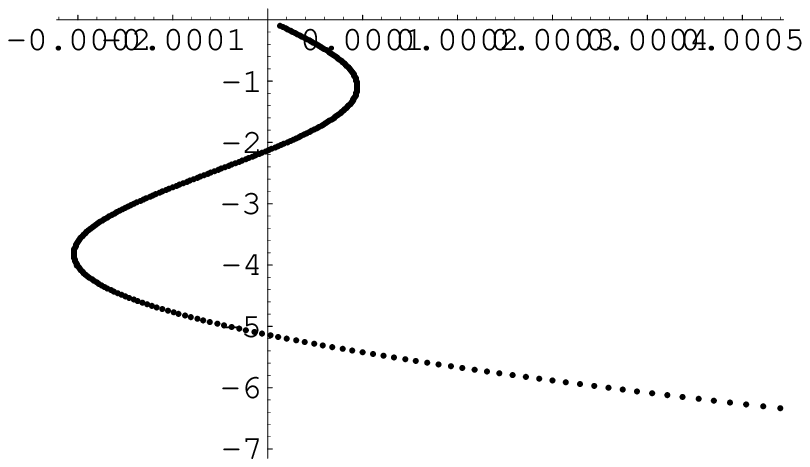}
\includegraphics[]{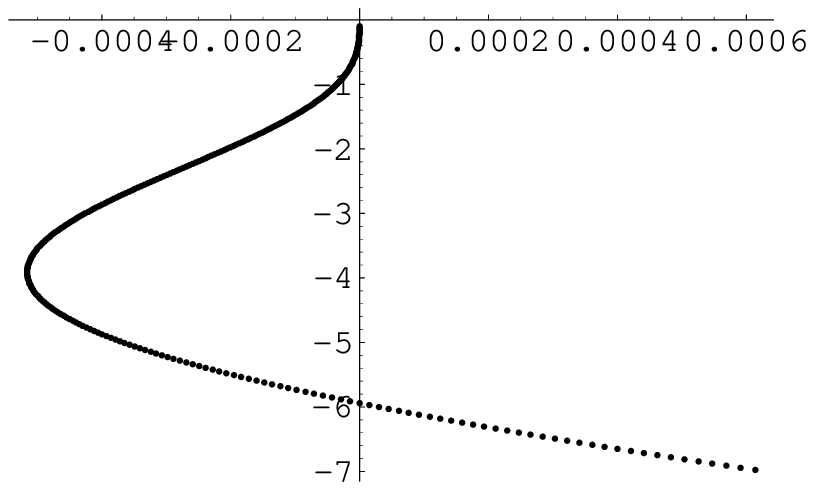}
\includegraphics[]{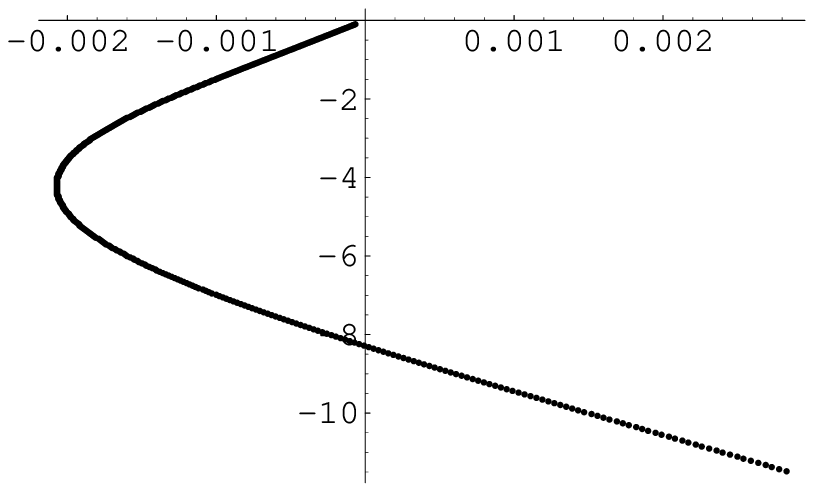}
\caption{The Tanaka's curves $\mcal{C}^+$ (i.e. $\bar{\Pi}^+$ versus $\bar{\Phi}^+$) in the theory (\ref{ipol})
with $g_4=300$. The
graphs are plotted, from the left to the right and from top to bottom,
for the values of $-V_M''/H_M^2=3.96,\ 3.98,\ 4.00$ and $4.10$ respectively.}
\label{tcfig}
\end{figure}
\begin{figure}
\centering
\includegraphics[]{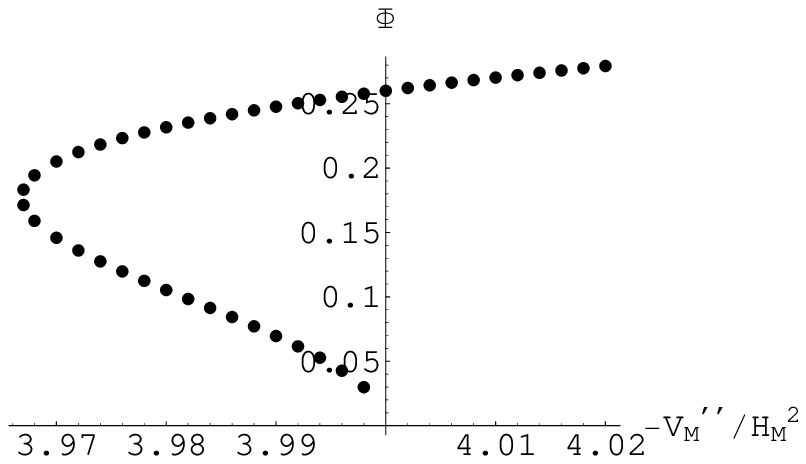}
\includegraphics[]{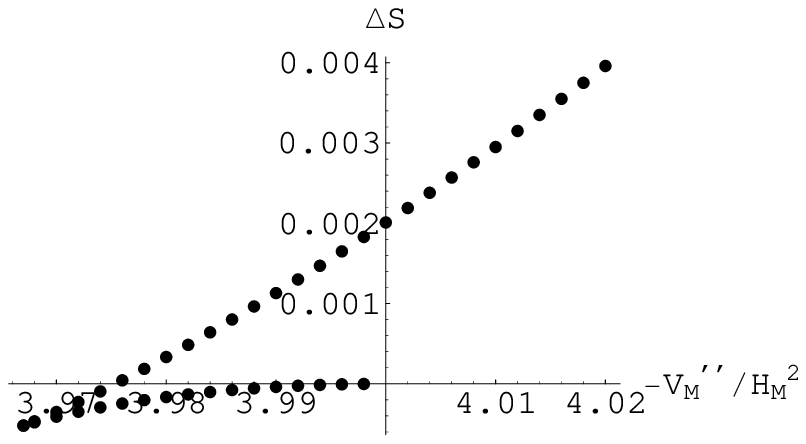}
\caption{Left graph: the initial value of the CdL instanton solution versus $-V_M''/H_M^2$ in the theory (\ref{ipol}) with
$g_4=300$.
There are two
bifurcation points in the parameter $-V_M''/H_M^2$ for solutions of
the instanton equations. At $-V_M''/H_M^2=4$ the number of
(nontrivial, i.e. no-HM-instanton) solutions changes from $1$ to $2$, and at a value approximately $3.966$ the
number of CdL instantons changes from $0$ to $2$. Right graph: the difference $\Delta S$ between the
action of CdL and HM intanton (the action of the HM instanton is normalized to $2$).
Lower curve describes the subcritical near-to-limit
CdL instanton and upper curve corresponds to the no-near-to-limit CdL instanton. These curves merge in the
point $-V_M''/H_M^2\approx 3.966$ mentioned above. For $-V_M''/H_M^2\gtrapprox 3.975$ the no-near-to-limit instanton governs
the vacuum decay, for the values below this the vacuum decay is governed by the HM instanton.
}
\label{instrfig}
\end{figure}

The existence of two CdL instantons of the
first order for given $-V_M''/H_M^2$ opens the question which instanton governs the vacuum decay. Let us
investigate a concrete realization of the kind of vacuum decay described above. In the appendix it is shown that
for a class of generalizations of the quartic potential we cannot obtain negative value of the fourth derivative of the
potential at $\Phi_M$ if we require that the potential contains both the false and true vacuum. Let us
relax these requirements and consider the potential
\be \label{ipol}
V(\Phi)=\frac{3}{8\pi}-\frac{1}{2}g_2\Phi^2-\frac{1}{24}g_4\Phi^4 \ ,
\ee
where $g_2,g_4$ are positive constants. The top of the potential is at $\Phi_M=0$ and
\begin{align*}
&V_M''=-g_2\ ,&
&H_M^2=1\ ,&
&-\frac{V_M''}{H_M^2}=g_2\ .&
\end{align*}
The choice
\bdis
g_4=300
\edis
ensures that
\bdis
8C+\frac{1}{48}\left(\frac{V_M'''}{H_M^2}\right)^2+\frac{1}{4}\frac{V_M''''}{H_M^2}=
8C-\frac{1}{4}g_4
\edis
is negative. We have performed numerical analysis of the structure of CdL instanton solutions in this theory for values
of $-V_M''/H_M^2$ close to $4$ from both sides. The structure of the instanton solution is fully characterized by the
Tanaka's curves $\mcal{C}$. Since the potential (\ref{ipol}) is an even function of $\Phi$ we need only one of the curves
$\mcal{C}^+$ and $\mcal{C}^-$ ($\mcal{C}^-$ is the mirror image of $\mcal{C}^+$ with respect to vertical axis in the
$(\bar{\Pi},\bar{\Phi})$ plane).
The instanton solutions are determined by the points at which $\mcal{C}^+$
(or $\mcal{C}^-$) crosses the vertical axis. The Tanaka's curves $\mcal{C}^+$ with $-V_M''/H_M^2$ close to $4$
and $g_4=300$ for the theory (\ref{ipol}) are shown on the graphs in figure \ref{tcfig}. These graphs tell us that
for $V_M''/H_M^2$ close above $4$ there is only one (no near-to-limit) CdL instanton, for $-V_M''/H_M^2=4$ lying
between approximately $3.966$ and $4$ there are two CdL instantons, and for $-V_M''/H_M^2$ less than approximately
$3.966$ there are no CdL instantons. Finally, the structure of the instanton solutions in the theory (\ref{ipol}) and
the $-V_M''/H_M^2$-dependence of the instanton action are
shown in figure \ref{instrfig}.

\section{Conclusion}

The false vacuum decay in a de Sitter universe has been investigated for near-to-critical values of the curvature
of the effective potential. An approximate formula for the Euclidean action of the near-to-limit CdL instanton has
been found by expanding the inflaton and the metric into to powers of the inflaton in a different way than in our
previous work \cite{vj2}. We have focused on the case when the fourth derivative
of the effective potential at its top has a subcritical value and $-V_M''/H_M^2$ is running from both sides
around its critical value $4$.
We conclude that there is a range of the parameter $-V_M''/H_M^2$ less than $4$ for which at least
two CdL instantons exist. One of them
is the near-to-limit instanton that can be described by the approximate formulas, together with its action,
derived in the first part of the paper.
The second instanton must exist because of the necessity to disconnect the energy curve from the potential when the
starting point of the curve moves towards the true vacuum \cite{vj1}.
The near-to-limit instanton in this case mediates the vaccum decay with
a less probability than the related HM instanton. However, the
vacuum decay is not governed by the HM instanton in this case but by the no-near-to-limit CdL instanton. On the
other hand, we have shown on a concrete example that for sufficiently small values of $-V_M''/H_M^2$ the HM
instanton has the least action from the three instantons in question and is to be considered as the instanton
governing the vacuum decay.

\noi
{\bf Acknowledgment}: I would like to thank Vlado Balek for fruitful discussions. This work was supported by the
Slovak grant VEGA 1/0250/03.

\appendix

\section{Non-negativeness of $V_M''''$ for a class of generalizations of quartic potential}

We will consider the following class of potentials
\be \label{polypot}
V_n=\frac{1}{2n}\Phi^{2n}-\frac{1}{2n+1}\delta\Phi^{2n+1}+\frac{1}{2n+2}\lambda\Phi^{2n+2}
\ee
with $n$ an arbitrary integer and $\lambda$ and $\delta$ are positive parameters. Requirement of non-negativeness of
$V_n$ and existence of both vacua leads to the restriction on $\delta$ at given $\lambda$
\bdis
\delta_m^{(n)}=2\sqrt{\lambda}<
\delta<\frac{2+\frac{1}{n}}{\sqrt{1+\frac{1}{n}}}\sqrt{\lambda}=\delta_M^{(n)} .
\edis
The width $\delta_M^{(n)}-\delta_m^{(n)}$ of the allowed interval of $\delta$ tends to zero as $n$ grows to infinity.
We want to answer the question whether it is possible to choose the parameters $\delta$ and $\lambda$ in such a way that
we obtain a negative denominator in the fraction on the right hand side of (\ref{k2quant}). Negativeness of
$V_M''''$ is obviously necessary for this. The top of the barrier of $V_n$ is reached at  $\Phi_M$ for which we have
\bdis
\delta\Phi_M=\frac{2}{1+Z}, \quad \mbox{with} \quad
Z=\sqrt{1-\frac{4\lambda}{\delta^2}} .
\edis
The range of the parameter $Z$ follows from the range of $\delta$, namely
\be \label{Zrestr}
Z\in\left[ 0,\frac{1}{2n+1}\right] .
\ee
The direct computation of $V''''_{n\ M}$ with the crucial help of the parametrization $(\lambda,Z)$ gives the result
\be \label{V_Mingeneralzed}
V''''_{n M}=\frac{6}{\lambda^{n-2}}\frac{(1-Z)^{n-2}}{(1+Z)^{n-2}}
\left[ -4n^2+4n-1+\frac{2n(2n-1)}{1+Z}\right] .
\ee
For any given $n$ there are obviously values of $Z$ (not fulfilling (\ref{Zrestr})) for which
$V''''_{nM}$ has both positive and negative values. The restriction (\ref{Zrestr}) changes
the situation. At $Z=0$ the expression (\ref{V_Mingeneralzed}) is positive. The zero point of (\ref{V_Mingeneralzed}) is
given by
\bdis
Z_0=\frac{(2n-1)2n}{4n^2+1-4n}-1\longrightarrow^{n\to\infty} 0 ,
\edis
and we easily find that $Z_0$ never belongs to the required range of $Z$ because
\bdis
Z_0-\frac{1}{2n+1}=\frac{4n-2}{(4n^2+1-4n)(2n+1)}>0 .
\edis
Therefore $V''''_{n\ M}$ cannot be negative.

\end{document}